\documentclass[reprint,superscriptaddress,aps,pra,floatfix,nofootinbib,bibnotes]{revtex4-1}
\usepackage[colorlinks=true,linkcolor=black,citecolor=blue, urlcolor=blue,bookmarks=false]{hyperref}
\hypersetup{breaklinks=true}
\usepackage[utf8]{inputenc}
\usepackage{natbib}
\usepackage{bm}
\usepackage{dcolumn}
\usepackage{array}
\usepackage{epsfig}
\usepackage{amsmath}
\usepackage{amssymb}
\usepackage{multirow}
\usepackage{graphicx,subcaption}
\usepackage{feynmf}
\usepackage{tensor}
\usepackage{comment}
\usepackage{enumitem}
\graphicspath{{./fig/}}
\usepackage[normalem]{ulem}  
\usepackage[dvips]{color} 

\renewcommand{\sout}{\bgroup \color{red} \ULdepth=-.5ex \ULset}

\newcommand{\vev}[1]{\langle{#1}\rangle}
\newcommand{\bigp}[1]{\Big({#1}\Big)}

\newcommand{\qv}{\vec{q}\,}
\newcommand{\OO}{\mathcal{O}}

\newcommand{\Logf}{\ln\frac{Q^2}{\mu^2}}
\newcommand{\rhon}{{\rho_N}}

\begin{document}
\title{The $\phi$ meson with finite momentum in a dense medium}

\author{HyungJoo Kim}
\email{hugokm0322@gmail.com}
\affiliation{Asia Pacific Center for Theoretical Physics, Pohang, Gyeongbuk 37673, Korea}
\author{Philipp Gubler}
\email{gubler@post.j-parc.jp}
\affiliation{Advanced Science Research Center, Japan Atomic Energy Agency, Tokai, Ibaraki 319-1195, Japan}
\date{\today}
\begin{abstract}
The dispersion relation of the $\phi$ meson in nuclear matter is studied in a QCD sum rule approach. 
In a dense medium, longitudinal and transverse modes of vector particles 
can have independently modified dispersion relations due to broken Lorentz invariance. 
Employing the full set of independent operators and corresponding Wilson coefficients
up to operator dimension-6, the $\phi$ meson QCD sum rules are analyzed with changing densities and momenta. 
The non-trivial momentum dependence of the $\phi$ meson mass is found to have opposite signs 
for the longitudinal and transverse modes. 
Specifically, the mass is reduced by 5 MeV for the longitudinal mode, while its increase amounts to 7 MeV for the transverse mode, both at a momentum scale 
of 1 GeV. 
In an experiment which does not distinguish between longitudinal and transverse polarizations, this could 
in principle be seen as two separated peaks at large momenta. 
Taking however broadening effects into account, the momentum dependence will most likely be seen as 
a small but positive effective mass shift and an increased effective width for non-zero momenta.  
\end{abstract}


\maketitle
\section{Introduction}
The study of light vector mesons in a dense medium can provide important insights to our 
understanding of the origin of hadron masses, which is closely related to the breaking 
of chiral symmetry in vacuum. Vector mesons are especially well suited for experimental measurements 
of in-medium effects, as they can decay into dileptons, which do not feel the strong 
interaction and are therefore less distorted by the presence of the nuclear medium 
compared to hadronic decay products. 
Initially, the $\rho$ meson mass shift in nuclear 
matter was considered to be a suitable probe for the restoration of chiral symmetry at finite 
density. With the help of QCD sum rules, it was (within certain approximations) possible 
to relate this mass shift to the reduction of the chiral condensate $\langle \bar{q} q \rangle$ ($q$ here 
stands for $u$ or $d$ quarks) 
at finite density \cite{Hatsuda:1991ez}. 
Such a mass shift was later reported in Ref.\,\cite{Naruki:2005kd} 
from measurements of dilepton spectra in 12 GeV pA reactions at the E325 experiment at KEK. 
It was, however, also realized that QCD sum rules in the $\rho$ meson channel can 
be satisfied equally well by both mass shifted and
broadened $\rho$ meson peaks \cite{Klingl:1997kf,Leupold:1997dg}, the latter being 
obtained for instance in hadronic effective theory calculations \cite{Rapp:1999ej,Post:2003hu}. 
The experimental findings of Ref.\,\cite{Nasseripour:2007aa} point to 
similar conclusions of a broadened $\rho$ meson without any mass shift. 

With the difficulty of drawing any definite conclusions about the behavior of the 
$\rho$ meson at finite density and its relationship to chiral symmetry, attention has 
turned to mesons with smaller widths such as the $\omega$ and the $\phi$. 
About the $\omega$, a lot of experimental work has been done in recent years. 
See for instance Ref.\,\cite{Metag:2017yuh} for a review. On the theoretical 
side, a suggestion was put forward by one of the present authors that the finite 
density behavior of the $\omega$ together with the axial-vector meson $f_1(1285)$ 
could serve as an indicator of the restoration of chiral symmetry in nuclear matter \cite{Gubler:2016djf}. 

In this work, we will however focus on the $\phi$ meson and its modification at finite 
density. 
Its behavior in nuclear matter has been studied over the years (see, for instance, Refs.\,\cite{Hatsuda:1991ez,Ko:1992tp,Asakawa:1994tp,Klingl:1997tm,Kampfer:2002pj,Cabrera:2002hc} 
for early theoretical works), but has recently received renewed interest 
especially due to new experimental results \cite{Ishikawa:2004id,Muto:2005za,Sakuma:2006xc,Wood:2010ei,Polyanskiy:2010tj,Hartmann:2012ia,PhysRevLett.123.022002} 
and near-future experiments, such as E16 at J-PARC \cite{Aoki:2015qla}. 
While the first QCD sum rule calculations \cite{Hatsuda:1991ez,Asakawa:1994tp} predicted a decreasing 
$\phi$ meson mass with increasing density, works based on hadronic effective field theories rather 
suggested a strong broadening, leading to 
an about an order of 
magnitude larger width than in vacuum \cite{Ko:1992tp,Klingl:1997tm}. 
These theoretical calculations have been revised in recent years, taking into account novel analysis methods 
and improved knowledge of QCD condensates for QCD sum rules  \cite{Gubler:2014pta}, and incorporating new
experimental constraints on hadronic interactions and considering higher order terms in 
effective theory calculations  \cite{Gubler:2015yna,Gubler:2016itj,Cabrera:2016rnc,Cabrera:2017agk,Cobos-Martinez:2017vtr,Cobos-Martinez:2017woo}. 
The qualitative conclusions of all these theoretical works have, however, remained the same. 

On the experimental side, the result of the E325 experiment at KEK \cite{Muto:2005za} deserves 
special attention, as it is so far the only one that was able to simultaneously constrain both the 
mass shift and width of the $\phi$ meson at normal nuclear matter density. By measuring the dilepton 
spectra in 12 GeV pA reactions with carbon and copper targets, the KEK-PS E325 Collaboration reported 
a negative mass shift of 35 MeV and a width increased by a factor of 3.6 compared to its vacuum value. 
It is furthermore worth mentioning the most recent experimental result of the HADES collaboration 
in Ref.\,\cite{PhysRevLett.123.022002}, where 
in $\pi^-$A reactions with A$\,=\,$C and A$\,=\,$W and an incident 
pion momentum of 1.7 GeV, the $\phi$ was observed to receive a significantly stronger absorption for 
the larger W target compared to C. This result provides further evidence for a strong broadening effect of the 
$\phi$ meson in nuclear matter, similar to earlier experiments by the LEPS Collaboration \cite{Ishikawa:2004id}, 
the CLAS Collaboration \cite{Wood:2010ei} and at COSY-ANKE \cite{Polyanskiy:2010tj}.

In this paper, we will specifically study 
the momentum dependence of the $\phi$ meson energy (e.g. its dispersion 
relation) at finite density. While the dispersion relation of any particle in vacuum is fixed by Lorentz symmetry, 
this is no longer the case in a medium, which serves as a special frame of reference. 
This can hence lead to a modified dispersion relation at finite density. 
The $\phi$ meson is in this context presently of particular interest, as its dispersion relation 
will be studied at the J-PARC E16 experiment, which will start running in 2020 \cite{Aoki:2015qla}. 

The main goal of this work is to determine the non-trivial dispersion relation of 
the $\phi$ meson and to make predictions for the E16 experiment at J-PARC. 
We furthermore study the longitudinal and transverse polarizations of the $\phi$, which 
are equal in vacuum, but can behave differently in nuclear matter. 
For this purpose, we make use of the QCD sum rule method, for which the effect of 
broken Lorentz invariance is encoded as expectation values of non-scalar QCD operators (see Refs.\,\cite{Lee:1997zta,Leupold:1998bt}). 
These expectation values always vanish in vacuum, but become finite in a hot or/and dense medium. 
After a separate QCD sum rule study of the longitudinal and transverse modes, we will 
discuss what effects the modified dispersion relations may have on future 
experimental dilepton measurements. 

This paper is organized as follows. In Section II, we give a brief description of the formalism 
of QCD sum rules, which is followed by a discussion of the used input parameters in Section III. 
Section IV is devoted to the detailed results obtained in this study and to a discussion of 
potential consequences for experimental measurements. 
The paper is summarized and concluded in Section V. 

\section{QCD sum rules with finite three-momentum}\label{formalism}
For the purpose of studying the $\phi$ meson in nuclear matter, let us first consider the  correlation function of 
the vector current with strange quarks $j_\mu(x)=\bar{s}(x)\gamma_\mu s(x)$,
\begin{align}
    \Pi_{\mu\nu}(\omega,\qv)=i\int d^4x e^{iq\cdot x}\vev{T\{j_\mu(x) j_\nu(0) \}}_{\rhon},\label{correlationftn}
\end{align}
where $\vev{\cdot}_{\rhon}$ represents the expectation value taken 
with respect to the nuclear matter ground state with
density $\rhon$. The nuclear medium is assumed to be at rest. 
In vacuum or in the $|\qv|\to0$ limit there is only one invariant function in this correlation function, i.e. $\Pi(\omega^2)=-\frac{1}{3\omega^2}\Pi^\mu_\mu$. For finite $\qv$ in nuclear matter, however, longitudinal and transverse polarization states are distinguishable and their distinct components are expressed by
\begin{align}
    \Pi_L(\omega^2,\qv^2)&=\frac{1}{\qv^2}\Pi_{00}, \label{eq:Lcorrelator} \\
    \Pi_T(\omega^2,\qv^2)&=-\frac{1}{2}\left( \frac{1}{q^2}\Pi^\mu_\mu+\Pi_L \right). \label{eq:Tcorrelator}
\end{align}
In the deep space-like $q^2$ region (e.g. $\omega\to i\infty$ with $|\qv|$ held fixed \cite{Cohen:1994wm}), these are calculable using the operator product expansion(OPE),
\begin{align}
    \Pi^{OPE}(\omega^2,\qv^2)=\sum_n C_n(\omega^2,\qv^2)\vev{O_n}_{\rhon}.
    \label{OPE}
\end{align}
In the time-like $q^2$ region, the imaginary part of the correlation function is expressed by the spectral function (at zero temperature),
\begin{align}
    \rho(\omega^2,\qv^2)=\frac{1}{\pi}\text{Im}\Pi(\omega^2,\qv^2).
\end{align}

For discussing the $\qv$ dependence, it is convenient to 
substitute $\omega^2$ by $Q^2=-q^2$ and to re-express the OPE in terms of $Q^2$ and $\qv^2$, i.e. $C_n(\omega^2,\qv^2)\to C_n(Q^2,\qv^2)$. 
After this substitution, it becomes more apparent that the $\qv$ dependence can be divided into two types: 
(i) trivial and (ii) non-trivial momentum dependence. 
The trivial dependence comes from the $\qv^2$ absorbed in $Q^2$, while the non-trivial one comes from 
terms in which $\qv^2$ cannot be absorbed in $Q^2$. Naturally, OPE terms which contain only the $Q^2$ dependence do not violate Lorentz symmetry and 
keep the ordinary form of the dispersion relation: $\omega^2=m_{\phi}^2+\qv^2$. 
This happens for scalar operators and their Wilson coefficients.
On the other hand, a non-trivial $\qv$ dependence appears in Wilson coefficients of non-scalar operators and play an important role in medium. 
They cause the $\phi$ meson to have not only a modified dispersion relation, $\omega^2=m_{\phi}^2(\qv^2)+\qv^2$, but also 
to have different longitudinal and transverse polarization modes. 

With the above change of variables, the remaining $\qv^2$ now represents only the non-trivial momentum dependence. The same substitution is also applicable to the spectral function side. Throughout the rest of this paper, we redefine it as $\rho(\omega^2,\qv^2) \to \rho(q^2,\qv^2)$ and employ the following simple ``pole + continuum'' ansatz,
\begin{align}
    \rho(q^2,\qv^2) \approx & \,\, \frac{f(\qv^2)}{4\pi^2}\delta(q^2-m_\phi^2(\qv^2))\nonumber\\
    &+\frac{1}{4\pi^2}\Big(1+\frac{\alpha_s}{\pi}\Big)\theta(q^2-s_0(\qv^2)).
    \label{ansatz}
\end{align}
Here, all non-trivial momentum dependence is assumed to be encoded in the three spectral parameters, $m_\phi(\qv^2)$, 
$f(\qv^2)$, and $s_0(\qv^2)$, which denote the mass of the $\phi$, the coupling strength between the $\phi$ and $j_\mu|0\rangle$, and the threshold parameter of 
the continuum, respectively. 
Let us give a few comments about the validity of the above ansatz. The 
$\delta$-function in the first term incorporates both trivial (hence, Lorentz invariant) momentum dependence, which is encoded in $q^2$ and the non-trivial 
momentum dependence, encoded in the momentum dependence of $m_{\phi}(\qv^2)$ and $f(\qv^2)$. 
As mentioned in the introduction, the $\phi$ meson is expected to get broadened in nuclear matter, which is ignored in Eq.\,(\ref{ansatz}). 
According to hadronic effective theory calculations and experimental measurements of the 
dilepton spectrum and the 
transparency ratio, the width increases 
roughly to values in the range of $\Gamma_{\phi} = 15 - 100$ MeV at normal nuclear matter density, which is at least an 
order of magnitude smaller than the $\phi$ meson mass. In contrast to the $\rho$ meson, 
the $\phi$ can therefore be identified as a single peak even in nuclear matter, which justifies to approximate it as 
a $\delta$-function. The second term describes the large energy limit of the spectral function, which should satisfy 
Lorentz symmetry and therefore does not explicitly depend on momentum. 
The continuum threshold, however, which 
approximately corresponds to the location of the first excited state or to a physical threshold, can depend on $\qv^2$. 
Again, the form of $\theta(q^2-s_0(\qv^2))$ ensures that both trivial and non-trivial momentum dependence is properly taken into account.

\begin{figure}
\centering
\vspace{0.5cm}
\includegraphics[width=0.65\linewidth]{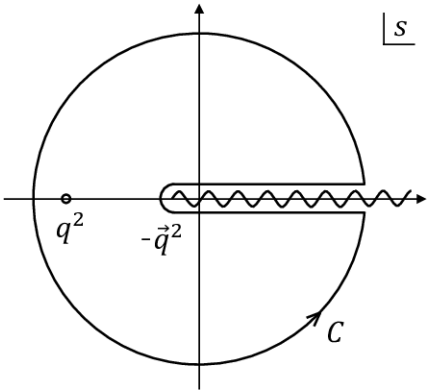}
\caption{The contour $\mathcal{C}$ on the complex plane of $s$, used in Eq.\,(\ref{Cauchy}) to derive the dispersion relation of Eq.\,(\ref{dispersion}).}
\label{contour}
\end{figure}

Using the variables $q^2$ and $\qv^2$, the standard dispersion relation which connects the spectral function to the OPE side can be 
derived by considering the contour integral illustrated in Fig.\,\ref{contour}.
Specifically, making use of the Cauchy theorem and the analyticity of $\Pi(q^2, \qv^2)$, we have 
\begin{align}
\Pi(q^2, \qv^2) = \frac{1}{2\pi i} \oint_{\mathcal{C}} ds \frac{\Pi(s, \qv^2)}{s - q^2}.
\label{Cauchy}
\end{align}
The integral along the outer circle of $\mathcal{C}$ can be assumed to vanish once its 
radius is taken to infinity (this assumption can for all practical cases be 
satisfied by introducing subtraction terms). Only the contour sections 
along the real axis of $s$ hence remain, which combined give rise to 
the imaginary part of $\Pi(q^2, \qv^2)$. 
One thus gets
\begin{align}
\Pi^{OPE}(Q^2,\qv^2)=\int^\infty_{-\qv^2} ds \frac{\rho(s,\qv^2)}{s+Q^2}. 
\label{dispersion}
\end{align}
We note that the above derivation is in essence the same as the one outlined in Ref.\,\cite{Leupold:1998bt}.

Eq.\,(\ref{dispersion}) is analogous to the vacuum dispersion relation except for the additional $\qv$ dependence. 
Therefore, we can apply the same analysis method as we do in 
the vacuum case. In other words, we can employ QCD sum rules for studying the $\phi$ meson in vacuum, at rest in medium, and in medium with finite 3-momentum within the 
same framework with changing density and 3-momentum. 
Note that this simplification is not applicable to baryons or charged mesons.

The conventional QCD sum rule analysis, which will be followed here, is based on the Borel transform. Given a general function $h(Q^2)$, it is defined as,
  \begin{align}
\widetilde{h}(M^2)\equiv\lim_{\substack{Q^2/n \rightarrow M^2, \\ n,Q^2
  \rightarrow \infty}}\frac{(Q^2)^n}{(n-1)!}\Big( -\frac{d}{dQ^2}\Big)^n h(Q^2).
\label{boreltransf}
\end{align}
where $M$ is called Borel mass. The analysis is performed through the following processes. For a given $\rhon$ and $|\qv|$, 
we can express the two spectral parameters, $m_\phi$ and $f$, as functions of $M^2$ and $s_0$,
\begin{align}
    m_\phi(M^2,s_0)&=\sqrt{M^2-\frac{\overline{\Pi}'(M^2,s_0)}{\overline{\Pi}(M^2,s_0)}},\\
    f(M^2,s_0)&=4\pi^2M^2\overline{\Pi}(M^2,s_0)e^{m_\phi(M^2,s_0)/M^2}.
\end{align}
Here, $\overline{\Pi}(M^2,s_0)\equiv\widetilde{\Pi}^{OPE}(M^2,\qv^2)-\frac{1}{M^2}\int^\infty_{s_0}ds e^{-s/M^2}\rho(s,\qv^2)$ and $\overline{\Pi}'(M^2,s_0)\equiv \partial \overline{\Pi}(M^2,s_0)/\partial(1/M^2)$. These functions are relevant only inside a so-called Borel window $(M_{min},M_{max})$, 
which is determined by the conditions that the pole contribution is larger than the continuum by 50$\%$ and that the dimension-6 contribution of the OPE series is smaller than $10\%$ of the total series. Specifically, we have
\begin{align}
 M_{max} \,&:   \frac{\int_{-\qv^2}^{s_0}dse^{-s/M^2}\rho(s,\qv^2)}{M^2\widetilde{\Pi}^{OPE}(M^2,\qv^2)}>0.5, \label{eq:Borel1} \\
  M_{min} \,&: \frac{\widetilde{\Pi}^{OPE}_{\mathrm{dim}\,6}(M^2,\qv^2)}{\widetilde{\Pi}^{OPE}(M^2,\qv^2)} <0.1. \label{eq:Borel2} 
\end{align}
We furthermore define average values of $m_\phi$ and $f$ 
inside this window as follows:
\begin{align}
    \overline{m}_\phi(s_0)&=\int^{M_{max}}_{M_{min}}dM \frac{m_\phi(M^2,s_0)}{M_{max}-M_{min}},\\
     \overline{f}(s_0)&=\int^{M_{max}}_{M_{min}}dM \frac{f(M^2,s_0)}{M_{max}-M_{min}}.
\end{align}
The next step is to find a threshold value $\overline{s}_0$ that makes the curve $m_\phi(M^2,s_0)$ the flattest. This is achieved by finding the minimum of $\chi^2_m$, defined as
\begin{align}
    \chi^2_m (s_0)=\int^{M_{max}}_{M_{min}}dM \frac{(m_\phi(M^2,s_0)-\overline{m}_\phi(s_0))^2}{M_{max}-M_{min}}.
\end{align}
Once $\overline{s}_0$ is determined, the other spectral parameters are obtained by their average values at this threshold value i.e. $\overline{m}_\phi(\overline{s}_0)$ and $\overline{f}_\phi(\overline{s}_0)$. 
The above process is repeated for all $\rhon$ and $|\qv|$ values to be investigated. 

\section{Input Parameters: Numerical Values and Uncertainties}
The complete set of operators relevant to the OPE of the vector channel up to dimension-6, discussed in Ref.~\cite{Kim:2017nyg}, is given by,
\begin{center}
    Scalar Operators
\begin{align}
    \bar{s}s,&   \label{Oss}\\
    G_0&\equiv \frac{\alpha_s}{\pi}G^a_{\mu\nu}G^a_{\mu\nu},\\
    \bar{s}js&\equiv g\bar{s}\gamma_{\mu}(D_{\nu}G_{\mu\nu})s, \\
    j^2&\equiv g^2(D_{\mu}G^a_{\alpha\mu})(D_{\nu}G^a_{\alpha\nu}),\\
    j_5^2 &\equiv g^2\bar{s} t^a \gamma_5 \gamma_{\mu}s \bar{s} t^a \gamma_5 \gamma_{\mu}s,
\end{align}
\end{center}
\begin{center}
    Non-Scalar(Twist) Operators
\begin{align}
A_{\alpha\beta}&\equiv g\mathcal{ST}\bar{s}(D_{\mu}G_{\alpha\mu})\gamma_{\beta}s, \\
B_{\alpha\beta}&\equiv g\mathcal{ST}\bar{s}\{iD_{\alpha},\tilde{G}_{\beta\mu}\}\gamma_{5}\gamma_{\mu}s, \\
C_{\alpha\beta}&\equiv m_s\mathcal{ST}\bar{s}D_{\alpha}D_{\beta}s, \\
F_{\alpha\beta}&\equiv\mathcal{ST}\bar{s}\gamma_{\alpha}iD_{\beta}s, \\
H_{\alpha\beta}&\equiv g^2\mathcal{ST}\bar{s} t^a \gamma_5 \gamma_{\alpha}s \bar{s} t^a \gamma_5 \gamma_{\beta}s, \\
K_{\alpha\beta\gamma\delta}&\equiv\mathcal{ST}\bar{s}\gamma_{\alpha}D_{\beta}D_{\gamma}D_{\delta}s, \\
G_{2\alpha\beta} &\equiv \frac{\alpha_s}{\pi} \mathcal{ST}G^a_{\alpha\mu}G^a_{\beta\mu},\\
X_{\alpha\beta}&\equiv\frac{\alpha_s}{\pi} \mathcal{ST}G^a_{\mu\nu}D_{\beta}D_{\alpha}G^a_{\mu\nu}, \label{eq:op1} \\
Y_{\alpha\beta}&\equiv\frac{\alpha_s}{\pi} \mathcal{ST}G^a_{\alpha\mu}D_{\mu}D_{\nu}G^a_{\beta\nu}, \label{eq:op2} \\
Z_{\alpha\beta}&\equiv\frac{\alpha_s}{\pi} \mathcal{ST}G^a_{\alpha\mu}D_{\beta}D_{\nu}G^a_{\mu\nu}, \label{eq:op3} \\
G_{4\alpha\beta\gamma\delta}&\equiv\frac{\alpha_s}{\pi}\mathcal{ST} G^a_{\alpha\mu}D_{\delta}D_{\gamma}G^a_{\beta\mu}.
\label{OG4}
\end{align}
\end{center}
$\mathcal{ST}$ here stands for the operation that makes the Lorentz indices symmetric and traceless. Non-scalar operators 
can also be categorized according to their twist (= dimension - spin). 

The expectation values of most of the above operators are not well known. Therefore, we often have to rely on assumptions and 
approximations which can give no more than order of magnitude estimates. In this section, we will succinctly discuss these techniques used to evaluate the various 
condensates appearing in this work.
The final values and uncertainties of all the input parameters are summarized in Table~\ref{inputparameters}. 

We will throughout this work make use of the linear density approximation, which can give a qualitatively good description to the 
condensates at the level of the normal nuclear matter density $\rho_0$. This approximation can be expressed as,
\begin{align}
    \vev{\mathcal{O}}_{\rhon} \approx\vev{\mathcal{O}}_0+\rhon\vev{\mathcal{O}}_N,
\end{align}
where $\vev{\OO}_0=\vev{0|\OO|0}$ and $\vev{\OO}_N=\vev{N(0)|\OO|N(0)}$, respectively. 
Here, $|N(\qv)\rangle$ is a one-nucleon state normalized as $\vev{N(\qv)|N(\qv')}=(2\pi)^3\delta^{(3)}(\qv-\qv')$. 
Lorentz indices of the nucleon matrix elements will be projected onto the nucleon four-momentum $p_\mu=(M_N,0,0,0)$. 
Therefore, throughout this paper, non-scalar condensates whose Lorentz indices are omitted, 
should be understood to be 
defined as $\vev{\OO_{\mu_1 \ldots \mu_2}}_N\equiv \OO\cdot\mathcal{ST}(p_{\mu_1}\ldots p_{\mu_2})$.

Under the above basic assumption, the condensates of the relevant operators are estimated as discussed below. 
\begin{enumerate}[leftmargin=*]
 \item Scalar matrix elements
 
The matrix elements of scalar operators have been frequently discussed in the literature 
(see for instance Ref.\,\cite{Gubler:2018ctz} and the references cited therein). 
We thus here only give the final expressions, which read

\begin{align}
    \vev{\bar{s}s}_{\rhon}&=0.8\vev{\bar{q}q}_0+\rhon\frac{\sigma_{sN}}{2m_s}, \label{eq:stangequarkcond} \\
    \vev{G_0}_{\rhon}&=\vev{G_0}_0+\rhon\frac{8}{9}(\sigma_{\pi N}+\sigma_{sN}-M_N), \label{eq:gluoncond}
    \end{align}
where $\sigma_{\pi N}=2m_q \vev{\bar{q}q}_N$ and $\sigma_{s N} = m_s \vev{\bar{s}s}_N$. The ratio $\vev{\bar{s}s}_0/\vev{\bar{q}q}_0\approx0.8$ is taken 
from an old QCDSR analysis \cite{Reinders:1984sr}. 
    For the scalar four-quark condensates, we have
    \begin{align}
    \vev{j^2}_{\rhon}&=(4\pi\alpha_s)^2\sum\vev{(\bar{q}\gamma_\mu t^aq)^2}_{\rhon} \simeq 0, \\
    \vev{j^2_5}_{\rhon}&=\frac{16}{36}4\pi\alpha_s\vev{\bar{s}s}^2_{\rhon},\\
    \vev{\bar{s}js}_{\rhon}&=-\frac{16}{36}4\pi\alpha_s\vev{\bar{s}s}^2_{\rhon}.
    \end{align}
The condensate on the first line is ignored here because it is proportional to $\alpha_s^2$ by use of the equation of motion. 
The others are estimated using the vacuum saturation approximation, which is assumed to hold at finite density in the same way as in vacuum. 

    \item Twist-2 matrix elements
   
   Matrix elements of twist-2 operators can be related to the moments of parton distribution functions measured in DIS experiments \cite{Bardeen:1978yd},
\begin{align}
    F&=\frac{1}{2M_N}A^s_2,\\
    K&=\frac{1}{2M_N}A^s_4,\\
    G_2&=-\frac{\alpha_s}{\pi}\frac{A^g_2}{M_N},\\
    G_4&=\frac{\alpha_s}{\pi}\frac{A^g_4}{M_N},
\end{align}
where $A^{s/g}_n$ is the $n$-th moment of the strange quark/gluon distribution function of the nucleon.
The values of $A_2^s$, $A_4^s$, $A_2^g$, and $A_4^g$ are listed in Table.~\ref{inputparameters}. 
They are extracted from the parton distributions provided in Ref.\,\cite{Harland-Lang:2014zoa} 
(see Ref.\,\cite{Gubler:2018ctz} for more details).

\item Quark twist-4 matrix elements

The $A$, $B$, $C$, and $H$ condensates were recently estimated in Ref.~\cite{Gubler:2015uza}, making use of 
the general assumption $\vev{\bar{s}\Gamma \OO s}_N \simeq \vev{\bar{q}\Gamma\OO q}_N\frac{A^s_2}{A^u_2}$. Here, 
we just show the final expressions, and refer the interested reader to Ref.\,\cite{Gubler:2015uza}. 
\begin{align}
    A&=\frac{1}{2M_N}K_u^2\frac{A^s_2}{A^u_2},\\
    B&=\frac{1}{2M_N}K_u^g\frac{A^s_2}{A^u_2},\\
    C&=-m_se^s_2\frac{A^s_2}{A^u_2},\\
    H&=\frac{1}{2M_N}(K^1_u-\frac{1}{2}K^1_{ud})\Big(\frac{A^s_2}{A^u_2}\Big)^2.
\end{align}
$e^s_2$ is the second moment of the twist-3 strange quark distribution function and $K^1_u, K^2_u, K^g_u$, 
and $K^1_{ud}$ denote matrix elements of up or down quark twist-4 operators.
Six different sets of $K_u^1$, $K_u^2$, and $K_u^g$ are given in Ref.~\cite{Choi:1993cu}. For their numerical values, 
we simply take the average over all the six sets for each parameter.
Their uncertainty is estimated as half of the difference between maximum and minimum of all sets.

    \item Gluon twist-4 matrix elements

The $X$, $Y$, and $Z$ condensates are generally not well known. 
To have a rough idea on their numerical values and systematic uncertainties,
we average two independent estimation methods and define the 
uncertainty as half of the difference between them. 
Specifically, we have
\begin{align}
    (X_1,Y_1,Z_1)&=\Big(\frac{1}{2M_N}\frac{0.32}{8.1}M_N M_N^0,0,\frac{3}{2}\frac{\alpha_s}{\pi}\frac{K^g_u}{2M_N} \Big),\\
    (X_2,Y_2,Z_2)&=\Big(\frac{-\vev{G_0}_N}{4},\frac{3G_2 M_N^2+\vev{G_0}_N}{48},3Y_2 \Big),
\end{align}
where $M_N^0$ denotes nucleon mass in the chiral limit \cite{Kim:2000kj}.
The first estimate, $(X_1,Y_1,Z_1)$, is taken from Ref.~\cite{Kim:2000kj}. The second one, $(X_2,Y_2,Z_2)$, is based on the method proposed in Ref.~\cite{Lee:1997zta},
\begin{align}
    \vev{ D_\mu G_{\rho\sigma}\ldots}_N&\approx -iP^g_\mu\vev{  G_{\rho\sigma}\ldots}_N\nonumber\\&\approx-i \frac{1}{2}p_\mu \vev{ G_{\rho\sigma}\ldots}_N.
\end{align}
Here, $P^g_\mu$ denotes the average momentum of the gluon in the nucleon, which is assumed to carry half of 
the nucleon momentum $p_\mu$.
In case of $Z$, however, the two estimates give quite similar values, so we just pick the first estimate because its 
uncertainty (which is related to $K^g_u$) is about 20 times larger than that of the second.
\end{enumerate}

\begin{table}[htbp]
\begin{center}
\caption{Input parameters, given at a renormalization scale of 1 GeV.} 
\label{inputparameters}
\begin{tabular}{ccc}  
\toprule
input parameter & value (uncertainty) & reference \\ 
\hline
$\alpha_s$& 0.472(0.024)\footnotemark[1] & \cite{Tanabashi:2018oca} \\
$m_s$ & 0.1242(0.0011) GeV\footnotemark[2]&\cite{Aoki:2019cca}  \\
$\rho_0$&0.17 $\text{fm}^{\text{-}3}$&\\
$M_N$ &(0.93827 + 0.93957)/2 GeV&\cite{Tanabashi:2018oca}\\
$M_N^0$&0.75 GeV&\cite{Kim:2000kj}\\
\hline
$\vev{\bar{q}q}_0$&-$(0.272(0.005))^3$/1.35 $\text{GeV}^3$\footnotemark[3]& \cite{Aoki:2019cca}\\
$\vev{G_0}_0$&0.012(0.004) $\text{GeV}^4$&\cite{Shifman:1978by,Shifman:1978bx}\\
$\sigma_{\pi N}$& 0.0397(0.0036) GeV&\cite{Aoki:2019cca}\\
$\sigma_{sN}$& 0.0529(0.007) GeV& \cite{Aoki:2019cca}\\
\hline
$A^u_2$&0.784(0.017) &\multirow{6}{*}{\cite{Gubler:2018ctz}}\\
$A^s_2$& 0.053(0.013)&\\
$A^g_2$& 0.367(0.023)&\\
$A^s_4$& 0.00121(0.00044)&\\
$A^g_4$& 0.0208(0.0023)&\\
$e^s_2$&0.00115(0.00318) &\\
\hline
$K_u^1$&0(0.173) $\text{GeV}^2$&\multirow{4}{*}{\cite{Choi:1993cu}}\\
$K_u^2$&-0.057(0.26) $\text{GeV}^2$&\\
$K_u^g$&-0.411(0.173) $\text{GeV}^2$&\\
$K_{ud}^1$&-0.083 $\text{GeV}^2$&\\
\hline
$X$&0.1014(0.0866) $\text{GeV}^4$&\\
$Y$&-0.0094(0.0094)  $\text{GeV}^4$&\\\botrule
\end{tabular}
\footnotetext[1]{Eq.~(9.4) in \cite{Tanabashi:2018oca} with $\Lambda_{\overline{MS}}^{(3)}=(332\pm17)$ MeV is used at the renormalization point $\mu$=1 GeV.}
\footnotetext[2]{Multiplied by 1.35 to rescale to $\mu$=1 GeV \cite{Tanabashi:2018oca}.}
\footnotetext[3]{Divided by 1.35 because $m_q\vev{\bar{q}q}$ is an RG invariant.}
\end{center}
\end{table}

\section{Results}
\subsection{Spectral parameters ($m_\phi$, $f$, $s_0$)}
\begin{figure}[h]
\centering
\includegraphics[width=0.7\linewidth]{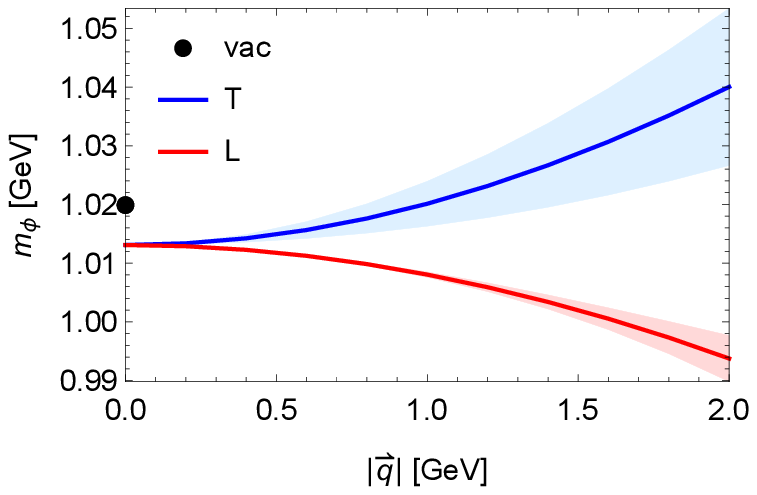}
\includegraphics[width=0.7\linewidth]{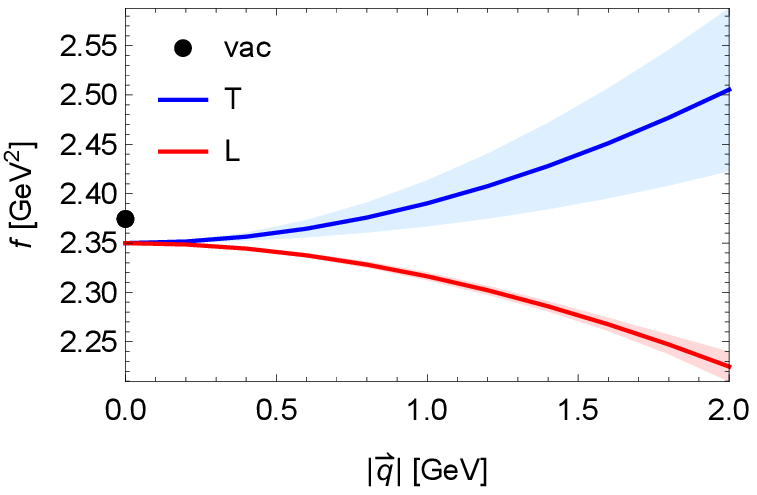}
\includegraphics[width=0.7\linewidth]{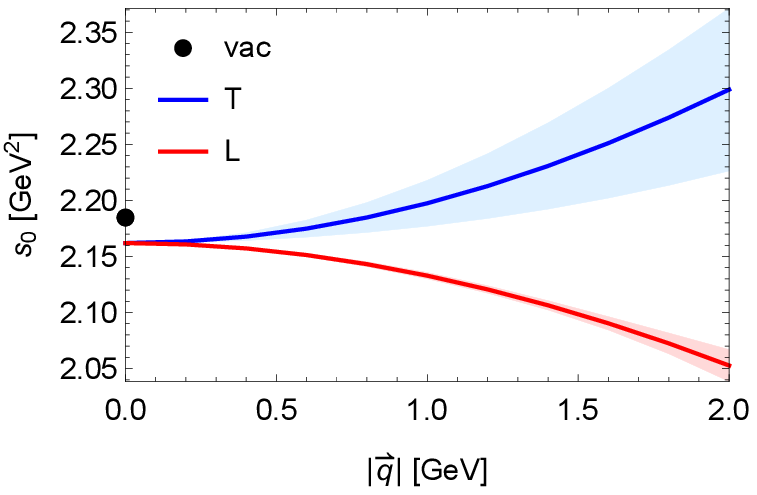}
\caption{Plot of $m_\phi$, $f$, and $s_0$ as a function of $|\qv|$ at normal nuclear matter density. The black dots denote the vacuum results. Blue (red) lines show the results for the transverse (longitudinal) mode.}
\label{mainplot}
\end{figure}

The main result of this work is the computed momentum dependence of the three spectral parameters
[$m_\phi(\qv^2)$, $f(\qv^2)$, $s_0(\qv^2)$] at normal nuclear matter density. 
In Fig.~\ref{mainplot} we plot these parameters up to 
$|\qv|=2.0$ GeV together with their vacuum values. The parameters exhibit a common behaviour 
at finite density. 
First, all of them are shifted negatively at zero momentum. 
This is caused primarily by the modification of the dimension-4 scalar condensates, 
shown in Eqs.\,(\ref{eq:stangequarkcond}) and 
(\ref{eq:gluoncond}), the magnitude of the shift being especially sensitive to the value of the strange 
sigma term $\sigma_{sN}$ \cite{Gubler:2014pta}.
The transverse mode parameters then increase, while the  longitudinal ones decrease with growing $|\qv|$.
The results shown in Fig.~\ref{mainplot} have rather large uncertainties coming from the errors 
of $m_s$, $\alpha_s$, $\vev{\bar{s}s}_0$ and $\vev{G_0}_0$, which determine the vacuum spectral parameters, 
and the modification of $\vev{\bar{s}s}_{\rhon}$ and $\vev{G_0}_{\rhon}$ at finite $\rhon$.
These uncertainties, however, only lead to an overall shift at zero momentum. Because our main 
interest is the momentum dependence of $m_\phi$ which is governed by the non-scalar condensates, 
we investigated the total uncertainty of $\Delta m_{\phi}(\qv^2)=m_{\phi}(\qv^2)-m_{\phi}(0)$ 
for which such an effect is reduced.
The error of all contributing parameters (here generally denoted as $\delta a_i$) 
is given in Table.~\ref{inputparameters}. The total uncertainty is 
estimated as $\sqrt{\sum_{i} (\Delta m_{\phi}|_{a_i \to a_i + \delta a_i} - \Delta m_{\phi})^2}$. 
For completeness, uncertainties of the 
other spectral parameters are computed in the same way and the results are shown as 
bands in Fig.~\ref{mainplot}. 
As can be observed there, 
the tendency of the momentum dependence is maintained even when taking into account the full error 
ranges of the various input parameters.

Both longitudinal and transverse modes exhibit an essentially quadratic behavior 
with respect to momentum $\qv$. Therefore, it is worthwhile to recast and parametrize the $m_\phi$-plot 
of Fig.~\ref{mainplot} into the following simple formula: 
\begin{align}
\frac{m_{\phi}^{\mathrm{L}/\mathrm{T}}(\rhon,\qv)}{m^{\text{vac}}_{\phi}} = 
1 + \bigg(a + b^{\mathrm{L}/\mathrm{T}} |\qv|^2 \bigg)\frac{\rhon}{\rho_0}.
\label{eq:simple_fit}
\end{align}
From our result at zero momentum, we have $m^{\text{vac}}_{\phi}=1.020\,\mathrm{GeV}$ and $a=-0.0067$, which strongly depends on the chosen value of $\sigma_{sN}$ (see Ref.\,\cite{Gubler:2014pta} for a detailed discussion). 
A fit to the dispersion relation curves then gives $b^\mathrm{T} = 0.0067 \pm 0.0034 \,\mathrm{GeV^{\text{-}2}}$ and $b^\mathrm{L} = -0.0048 \pm 0.0008\,\mathrm{GeV}^{\text{-}2}$. 
Here, we note that the full momentum dependence
of each OPE term is taken into account in our calculation. 
Therefore, $|\qv|$ can in principle be increased until the OPE convergence breaks down. 
We have checked that Eq.(\ref{eq:simple_fit}) is valid up to 
$|\qv| \simeq 3.0$\,GeV, above which the sum rule analysis results start to deviate from the quadratic behavior. 
Moreover, for $|\qv| \gtrsim 4.5$\,GeV, a valid $M_{min}$ can no longer be defined according to the condition of Eq.(\ref{eq:Borel2}).

\begin{figure}[h]
\centering
\includegraphics[width=0.7\linewidth]{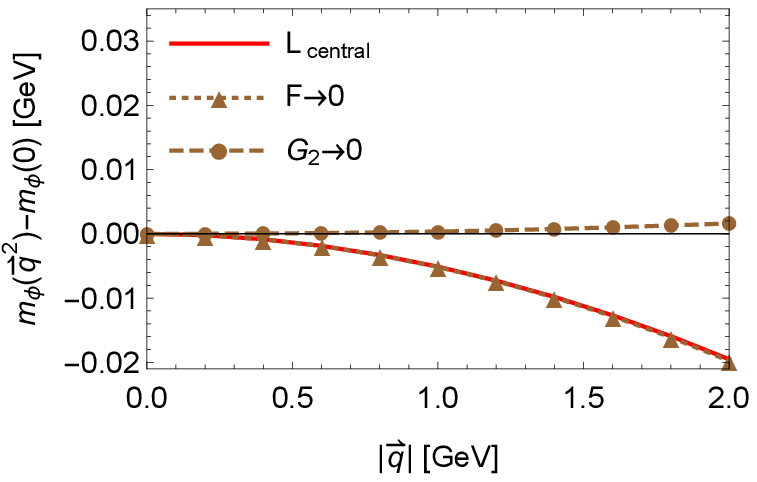}
\includegraphics[width=0.7\linewidth]{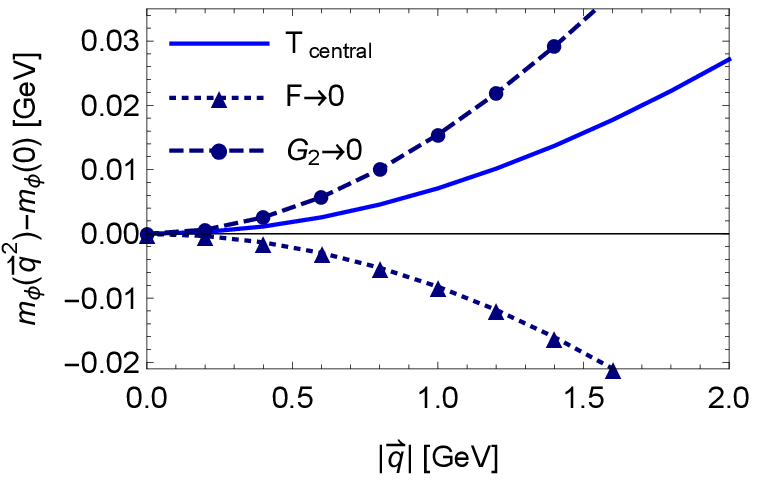}
\caption{The effects of the non-scalar condensates $F$ and $G_2$ on $\Delta m_{\phi}(\qv^2)$. The upper (lower) plot shows the behavior of the longitudinal (transverse) mode.}
\label{errorplot}
\end{figure}

Furthermore, we investigated which condensates primarily determine the 3-momentum dependence
of $\Delta m_{\phi}(\qv^2)$ for both polarization modes. 
For this purpose, we simply set each condensate to zero and compute how $\Delta m_{\phi}(\qv^2)$ changes. 
As a result, we found that only the two dimension-4 twist-2 condensates, 
i.e. $F$ and $G_2$, significantly change the momentum dependence as illustrated in Fig.~\ref{errorplot}. 
It can be seen in this figure that, for the longitudinal mode, 
the $F$ term has practically no effect while the 
negative mass shift is almost completely caused by the $G_2$ term. 
The behavior of the transverse mode, on the other hand, results from a large positive contribution from the 
$F$ term, which is reduced somewhat by the $G_2$ term.
All other condensates only have minor effects below $|\qv|=2$ GeV. 

\subsection{Unpolarized spectra with finite widths}
\begin{figure}[h]
\centering
\includegraphics[width=1.0\linewidth]{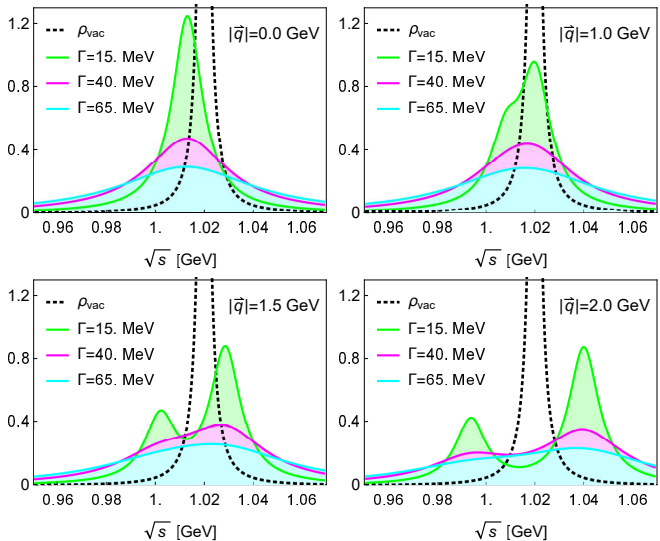}
\caption{The polarization-averaged $\phi$ meson peak with $\Gamma$=15, 40, and 65 MeV at normal nuclear matter density. The vacuum peak is shown as a black dotted line for comparison.}
\label{singleBW}
\end{figure}

In most experiments, dilepton measurements have so far not distinguished the two polarization modes. 
Thus, in general, the sum of both modes will be measured together with the vacuum peak in unpolarized
spectra\footnote{In an actual experiment such as that reported in Ref.\,\cite{Muto:2005za}, $\phi$ mesons are 
produced in nuclei with a finite size. A large fraction of them decay into dileptons only after they have moved outside 
of the dense nucleon region. Such dileptons will only contribute 
as vacuum peaks to the observed spectrum.}. 
Moreover, even though we have assumed the $\phi$ meson width to be 
zero in our analysis, this is not the case in reality (see for instance Refs.\,\cite{Gubler:2015yna,Cabrera:2016rnc}). Thus, to have a more realistic idea about the behavior of the dilepton spectrum at normal 
nuclear matter density and non-zero momentum, we artificially introduce a width using the relativistic Breit-Wigner form, 
\begin{align}
 \rho_{{}_\mathcal{ B}}(s,\qv^2)=\frac{f(\qv^2)}{4\pi^2}\frac{\sqrt{s}\,\Gamma/\pi}{(s-m^2_\phi(\qv^2))^2+s\Gamma^2},
\end{align}
where the values of $m_\phi(\qv^2)$ and $f(\qv^2)$ are taken from our results. The in-medium width of the $\phi$ meson is 
presently still not well known. 
We hence for illustration have tested three different values, 
$\Gamma$=15, 40, and 65 MeV. 
Among these, $\Gamma$=15 MeV represents the one reported by the 
E325 experiment in Ref.\,\cite{Muto:2005za}. The other two are representative 
for values obtained in hadronic effective theories \cite{Gubler:2015yna,Gubler:2016itj,Cabrera:2016rnc,Cabrera:2017agk,Cobos-Martinez:2017vtr,Cobos-Martinez:2017woo} and other experimental measurements \cite{Ishikawa:2004id,Wood:2010ei,Polyanskiy:2010tj,PhysRevLett.123.022002}. 
The polarization-averaged peak, 
$\frac{1}{3}(\rho^L_{{}_\mathcal{B}}(s,\qv^2)+2\rho^T_{{}_\mathcal{B}}(s,\qv^2))$ with these width values are shown in Fig.~\ref{singleBW} together with the vacuum peak ($\Gamma_{\text{vac}}$=4.26 MeV) for selected momenta. 

\begin{figure}[h]
\centering
\includegraphics[width=0.67\linewidth]{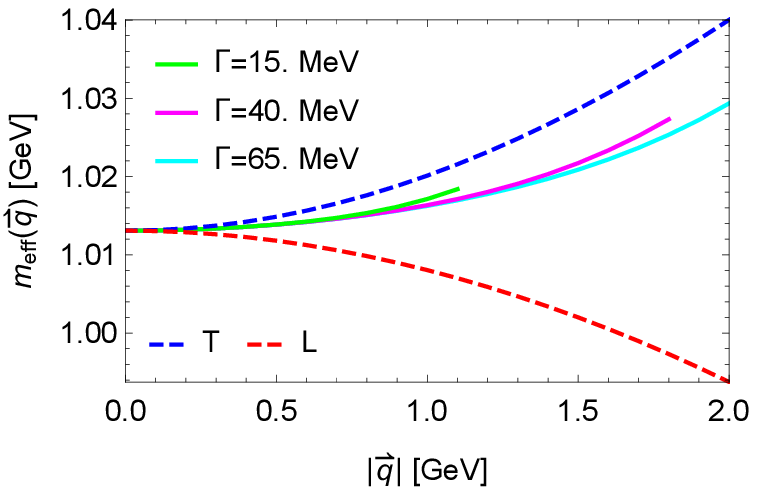}
\hspace*{0.6em}{\includegraphics[width=0.65\linewidth]{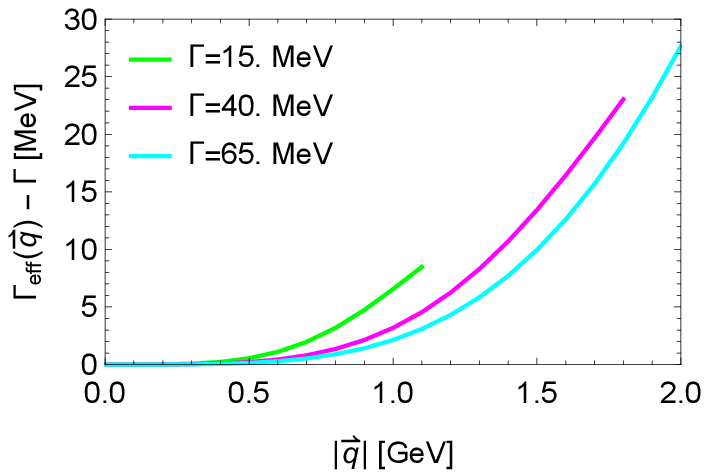}}
\caption{Effective mass (upper plot) and width increase (lower plot) of the single peak fit, shown as a function of $|\qv|$. In the upper plot, the central values of the transverse (longitudinal) masses are shown as blue (red) dashed lines for comparison.}
\label{avgmasswidth}
\end{figure}

We then fit the polarization-averaged peak using a single Breit-Wigner peak.
The effective mass and width extracted from the fit are plotted as a function of $|\qv|$ in Fig.~\ref{avgmasswidth}. They are shown as long as transverse and longitudinal peaks are separated less than their width.
As can be observed there, the mass shifts of the two modes partially cancel each other. Because there are two transverse and only one longitudinal mode, 
the average mass however tends to the transverse side and is thus positive. 
Furthermore, as the two components move away from each other, this leads 
to an increasing width with increasing momentum. 
Meanwhile, we found that generally the momentum dependence of the averaged peak is reduced as the in-medium width increases.

\subsection{Discussion}
Let us discuss potential implications of the obtained results for future dilepton measurements in the $\phi$ meson mass region. As can be seen in Fig.\,\ref{singleBW}, 
due to the opposite behavior of the longitudinal and transverse modes, 
the (angle averaged) dilepton spectrum can develop a double peak structure.
Although larger in-medium widths make it more difficult to observe such a double peak, it could in principle be seen at sufficiently 
large momenta.
Care is however needed when applying this result to experimental dilepton spectra, 
as $\phi$ mesons with large momenta likely travel outside of the nucleus before they decay 
into dileptons, which are therefore not strongly affected by the dense medium. 
Indeed, in the E325 experiment a finite density effect was only observed for dileptons with 
$\beta \gamma$ ($= |\qv|/m$) values smaller than 1.25 \cite{Muto:2005za}.
We therefore expect the average increase of the mass (green 
, pink and light blue curves 
of the upper plot in 
Fig.\,\ref{avgmasswidth}) and the increase of the width (lower plot in Fig.\,\ref{avgmasswidth}) 
to be the most easily and likely detectable finite momentum effects. 

We, however, stress that for independently measured longitudinal and transverse spectra, 
the mass shifts are larger. It would hence be interesting to measure not only the angle averaged 
dileptons but also their angular distributions, which would make it possible to disentangle 
the two contributions and to potentially observe the splitting between the two modes \cite{Bratkovskaya:1995kh,Speranza:2016tcg}. 
Specifically, the angular distribution of the positive (or negative) lepton in the rest frame of the dilepton pair is, after choosing a suitable polarization axis and averaging over the 
azimuthal angle, obtained as \cite{Speranza:2016tcg}
\begin{align}
\frac{dN}{d(\cos \theta)} & \propto \Sigma_{T}( 1 +  \cos^2 \theta) 
+ \Sigma_{L}( 1 - \cos^2 \theta) \nonumber \\
& \propto 1 + \lambda \cos^2 \theta, 
\label{angular_distribution}
\end{align}
with 
\begin{align}
\lambda = \frac{\Sigma_{T} - \Sigma_{L}}{\Sigma_{T} + \Sigma_{L}}. 
\label{angular_distribution_2}
\end{align}
Here, $\theta$ is the angle between the momentum of the lepton and 
the chosen polarization axis, while 
$\Sigma_{T}$ ($\Sigma_{L}$) are the contributions from the transversely (longitudinally) polarized vector mesons (or other dilepton sources). 
See, for instance, Ref.\,\cite{Abt:2009nu} for a discussion of 
commonly used polarization axes
and a measurement of $\lambda$ from $J/\psi$'s at 920 GeV pA collisions at the HERA-B experiment.

\section{Summary and Conclusions}
In this paper, we have analyzed the strange quark vector channel at finite density and momentum in a QCD sum rule approach. 
This was done with the goal of studying the dispersion relation of the 
longitudinal and transverse modes of the $\phi$ meson in nuclear medium. A linear combination of them will be measured at the E16 experiment at J-PARC. 
We found that both longitudinal and transverse dispersion relations are modified significantly compared with their vacuum form. 
At a momentum scale of 1 GeV, 
the longitudinal $\phi$ meson mass is reduced by 5 MeV, while its transverse counterpart is increased by 7 MeV. 
With a simultaneously increasing width, it is however unlikely that the two modes will be seen as separate peaks. 
Instead, their effect might only be detectable as an increasing effective mass and width with increasing momentum. 
Assuming the width to be fixed at 15 MeV in nuclear matter, 
we estimate the mass shift of the combined (one longitudinal and two transverse modes) $\phi$ meson peak at a momentum of 1 GeV to be 
about 4 MeV, while the width can be expected to increase by about 7 MeV at the same 
momentum. It remains to be seen whether e.g. the E16 experiment will have a sufficient resolution to detect these effects. 

\section*{Acknowledgements}
The authors thank Su Houng Lee for fruitful discussions and collaboration at 
the early stage of this project. 
P.G. is supported by Grant-in-Aid
for Early-Carrier Scientists No. JP18K13542 and the Leading Initiative for
Excellent Young Researchers (LEADER)
of the Japan Society for the Promotion of Science (JSPS). 
H.K. was supported by the YST Program at the APCTP through the Science and Technology Promotion Fund and Lottery Fund of the Korean Government and also by the Korean Local Governments - Gyeongsangbuk-do Province and Pohang City.

\section*{Appendix}
In this appendix, we give the principle formulas needed for the numerical analyses of this paper. They can be obtained by substituting the OPE results given in Ref.\,\cite{Kim:2017nyg} into Eqs.(\ref{eq:Lcorrelator}) and (\ref{eq:Tcorrelator}). For non-scalar operators, $\hat{\mathcal{O}}$ denotes $\mathcal{O} \times (M_N)^s$  where $s$ is spin of the operator.

\begin{widetext}
\begin{align}
\Pi_L(Q^2,\qv^2)&=-\frac{3m_s^2}{2\pi^2Q^2}-\frac{1+\alpha_s/\pi}{4\pi^2}\Logf+\Big(2-\frac{8m_s^2}{3Q^2}\Big)\frac{m_s\vev{\bar{s}s}}{Q^4}+\bigp{\frac{1}{12}+\frac{m_s^2}{9Q^2}}\frac{\vev{G^2}}{Q^4}+\bigp{\frac{1}{81}+\frac{2}{27}\Logf}\frac{\vev{j^2}}{Q^6} -\frac{2\vev{j_5^2}}{Q^6} \nonumber\\
&-\frac{4\vev{\bar{s}js}}{9Q^6}+\bigp{2-\frac{3m_s^2}{Q^2}}\frac{\hat{F}}{Q^4}-\frac{10\hat{K}}{3Q^6}+\bigp{\frac{3}{4}-\frac{11m_s^2}{12Q^2}-\bigp{\frac{1}{3}+\frac{4m_s^2}{3Q^2}}\Logf}\frac{\hat{G}_2}{Q^4}+\bigp{\frac{205}{216}-\frac{11}{36}\Logf}\frac{\hat{G}_4}{Q^6}\nonumber\\
&+\frac{\hat{A}}{2Q^6}-\frac{5\hat{B}}{2Q^6}+\frac{7\hat{C}}{Q^6}+\frac{2\hat{H}}{Q^6}+\bigp{\frac{1}{6}+\frac{5}{24}\Logf}\frac{\hat{X}}{Q^6}-\bigp{\frac{49}{144}+\frac{1}{8}\Logf}\frac{\hat{Y}}{Q^6}+\bigp{\frac{11}{16}+\frac{5}{8}\Logf}\frac{\hat{Z}}{Q^6} \nonumber\\
&+\frac{\qv^2}{Q^2}\Big\{\frac{\hat{A}}{Q^6}+\frac{3\hat{B}}{Q^6}-\frac{18\hat{C}}{Q^6}-\frac{6m_s^2\hat{F}}{Q^6}+\frac{4\hat{K}}{Q^6}-\bigp{\frac{2}{3}+\frac{11m_s^2}{6Q^2}-\frac{4m_s^2}{Q^2}\Logf}\frac{\hat{G}_2}{Q^4}-\bigp{\frac{373}{180}-\frac{11}{30}\Logf}\frac{\hat{G}_4}{Q^6}\nonumber\\
&-\bigp{\frac{4}{15}+\frac{1}{4}\Logf}\frac{\hat{X}}{Q^6}+\bigp{\frac{43}{360}+\frac{5}{12}\Logf}\frac{\hat{Y}}{Q^6}-\bigp{\frac{91}{120}+\frac{3}{4}\Logf}\frac{\hat{Z}}{Q^6} \Big\}+\frac{\qv^4}{Q^4}\Big\{\frac{16}{15}\frac{\hat{G}_4}{Q^6}\Big\},\\
\Pi_T(Q^2,\qv^2)&=\Pi_L(Q^2,0)+\frac{\qv^2}{Q^2}\Big\{-\bigp{4-\frac{9m_s^2}{Q^2}}\frac{\hat{F}}{Q^4}+\frac{18\hat{K}}{Q^6}-\bigp{\frac{7}{6}-\frac{11m_s^2}{4Q^2}-\bigp{\frac{2}{3}+\frac{2m_s^2}{3Q^2}}\Logf}\frac{\hat{G}_2}{Q^4}\nonumber\\
&-\bigp{\frac{559}{120}-\frac{33}{20}\Logf}\frac{\hat{G}_4}{Q^6}-\frac{3\hat{A}}{2Q^6}+\frac{7\hat{B}}{2Q^6}-\frac{5\hat{C}}{Q^6}-\frac{4\hat{H}}{Q^6}-\bigp{\frac{1}{5}+\frac{7}{24}\Logf}\frac{\hat{X}}{Q^6}\nonumber\\
&+\bigp{\frac{149}{240}+\frac{1}{24}\Logf}\frac{\hat{Y}}{Q^6}-\bigp{\frac{239}{240}+\frac{7}{8}\Logf}\frac{\hat{Z}}{Q^6}\Big\}+\frac{\qv^4}{Q^4}\Big\{\bigp{\frac{181}{45}-\frac{22}{15}\Logf}\frac{\hat{G}_4}{Q^6}-\frac{16\hat{K}}{Q^6}\Big\}.
\end{align}
\end{widetext}

\bibliographystyle{apsrev4-1}
\bibliography{phimesonbib}

\end{document}